\documentclass[%
 aip,
 jmp,
 reprint,%
]{revtex4-1}

\usepackage{graphicx}
\usepackage{dcolumn}
\usepackage{bm}

\usepackage[utf8]{inputenc}
\usepackage[T1]{fontenc}
\usepackage{mathptmx}
\usepackage{etoolbox}
\usepackage{CJKutf8}
\usepackage[version=3]{mhchem}
\usepackage[separate-uncertainty = true,multi-part-units=single]{siunitx}
\usepackage{listings}
\usepackage{csquotes}
\usepackage{gensymb}
\usepackage[dvipsnames]{xcolor}

\usepackage[colorlinks=true,
            linkcolor=blue,
            citecolor=blue,
            urlcolor=blue]{hyperref}

\begin{document}
\preprint{AIP/123-QED}

\title{Optimizing adsorption configurations on alloy surfaces using Tensor Train Optimizer}

\author{Tuan Minh Do}
\email{do.tuan.minh.qiqb@osaka-u.ac.jp}
\affiliation{Center for Quantum Information and Quantum Biology, The University of Osaka, 1-2 Machikaneyama, Toyonaka, Osaka 560-8531, Japan.}
\author{Tomoya Shiota}%
\affiliation{Center for Quantum Information and Quantum Biology, The University of Osaka, 1-2 Machikaneyama, Toyonaka, Osaka 560-8531, Japan.}
\affiliation{Graduate School of Engineering Science, The University of Osaka, 1-3 Machikaneyama, Toyonaka, Osaka 560-8531, Japan.}%
\author{Wataru Mizukami}
\email{mizukami.wataru.qiqb@osaka-u.ac.jp}
\affiliation{Center for Quantum Information and Quantum Biology, The University of Osaka, 1-2 Machikaneyama, Toyonaka, Osaka 560-8531, Japan.}
\affiliation{Graduate School of Engineering Science, The University of Osaka, 1-3 Machikaneyama, Toyonaka, Osaka 560-8531, Japan.}%

\date{\today}

\begin{abstract}
Understanding how molecules arrange on surfaces is fundamental to surface chemistry and essential for the rational design of catalytic and functional materials. In particular, the energetically most stable configuration provides valuable insight into adsorption-related processes. However, the search for this configuration is a global optimization problem with exponentially growing complexity as the number of adsorbates and possible adsorption sites increases. To address this, we express the adsorption energy as a sum of multi-adsorbate interaction terms, evaluated using our in-house trained machine learning interatomic potential MACE-Osaka24, and formulate the search for the most stable configuration as a higher-order unconstrained binary optimization (HUBO) problem. We employ a tensor-train-based method, Tensor Train Optimizer (TTOpt), to solve the HUBO problem and identify optimal adsorption configurations of CO and NO molecules on various alloys up to full surface coverage. Our results show that including interaction terms up to third order may be sufficient to approximate adsorption energies within chemical accuracy and to identify optimal configurations. We also observed that TTOpt performs better with the HUBO formulation, suggesting that third-order terms help preserve correlations between adsorption sites, which allow TTOpt to optimize configurations more effectively. The extensive benchmarks across various alloys, surface geometries, and adsorbates demonstrate the robustness and applicability of using TTOpt to solve HUBO-type global optimization problems in surface chemistry. In contrast to quantum and digital annealers, which have recently been applied to similar global optimization tasks but are restricted to cost functions with at most quadratic terms, our approach can incorporate higher-order terms in a straightforward manner and does not require specialized hardware.
\end{abstract}

\maketitle

\section{Introduction} \label{Introduction}
An atomistic understanding of adsorption is of fundamental importance in surface chemistry and heterogeneous catalysis, given that adsorption is one of the elementary steps that determine surface reactivity.\cite{somorjaiIntroductionSurfaceChemistry2010,norskovFundamentalConceptsHeterogeneous2014,koperThermodynamicTheoryMultielectron2011,liHeterostructuredElectrocatalystsFundamental2025} The adsorption of diatomic molecules such as CO and NO on solid surfaces is of particular interest, as they are frequently involved as reactants, intermediates, or products in chemical reactions relevant to key industrial applications, including CO and NO oxidation in automotive catalytic converters.\cite{somorjaiIntroductionSurfaceChemistry2010,norskovFundamentalConceptsHeterogeneous2014,koperThermodynamicTheoryMultielectron2011,liHeterostructuredElectrocatalystsFundamental2025} Their adsorption properties depend not only on adsorbate-surface interactions but also on competing adsorbate-adsorbate interactions, which makes accurate atomistic methods essential for analyzing the underlying processes.\cite{schmidtBenchmarkDatabaseTransition2018} Electronic structure calculations can provide such detailed insights, with density functional theory (DFT) being particularly popular for striking a good balance between computational efficiency and accuracy.\cite{kohnSelfConsistentEquationsIncluding1965,araujoAdsorptionEnergiesTransition2022} However, DFT calculations remain computationally expensive, especially for large-scale systems or high-throughput screening. Recently, machine learning interatomic potentials (MLIPs) have emerged as a promising alternative, offering first-principles accuracy at a fraction of the computational cost. \cite{behlerGeneralizedNeuralNetworkRepresentation2007,batatiaMACEHigherOrder2022,musaelianLearningLocalEquivariant2023,shiotaTamingMultiDomainFidelity2024}

The global minimum adsorption configuration is of particular interest as it plays a central role in approximate methods for predictive catalysis, including cluster expansion methods and descriptor-based screening approaches.\cite{tangLateralInteractionsOxygen2004,hanSurfaceSegregationOrdering2005,schmidtPerformanceClusterExpansions2012,zhangAtomisticUnderstandingHydrogen2025, jungMachinelearningDrivenGlobal2023,lanAdsorbMLLeapEfficiency2023,chenMultimodalTransformerPredicting2025}
However, identifying this configuration poses a substantial challenge, as it represents a high-dimensional global optimization problem with a discontinuous objective function. This makes gradient-based approaches unsuitable, while the search space increases exponentially with the number of possible adsorption sites and adsorbates. Even with MLIPs, exhaustive searches are only feasible for small systems, since the number of required calculations grows exponentially with system size.\cite{sampeiQuantumAnnealingBoosts2023} 
Consequently, there is a need for methods that can identify optimal adsorption configurations with sufficient accuracy by evaluating only a fraction of all possible configurations.

In recent years, there has been growing interest in quantum computing due to its ability to efficiently handle specific problems with exponentially increasing complexity.\cite{nielsenQuantumComputationQuantum2010,aruteQuantumSupremacyUsing2019} By leveraging quantum principles such as superposition and entanglement, quantum algorithms can explore multiple solutions simultaneously.\cite{nielsenQuantumComputationQuantum2010,aruteQuantumSupremacyUsing2019} One quantum computing technique that has matured to the point of commercial availability is quantum annealing\cite{brunoapolloniNumericalImplementationQuantum1990,kadowakiQuantumAnnealingTransverse1998,farhiQuantumAdiabaticEvolution2001, johnsonQuantumAnnealingManufactured2011,boixoExperimentalSignatureProgrammable2013, boixoEvidenceQuantumAnnealing2014,lechnerQuantumAnnealingArchitecture2015,haukePerspectivesQuantumAnnealing2020}, which is particularly well suited for optimization problems that can be formulated using the Ising model or, equivalently, expressed as quadratic unconstrained binary optimization (QUBO) problems. This progress has also sparked a parallel development of quantum-inspired algorithms---algorithms that run on classical computers but mimic principles of quantum computing. For example, digital annealers\cite{aramonPhysicsInspiredOptimizationQuadratic2019,fujisakiPracticalScalableDecoder2022, fukushima-kimuraMathematicalAspectsDigital2023,choubisaAcceleratedChemicalSpace2023} are inspired by quantum annealing and have been used to predict optimal adsorption configurations of CO molecules on a Pd-Zn surface.\cite{sampeiQuantumAnnealingBoosts2023} It has been reported that speed-ups of 10 to 50 times can be achieved by using digital annealers compared to conventional methods.\cite{choubisaAcceleratedChemicalSpace2023}

However, both quantum and digital annealing require specialized hardware, which restricts their accessibility to a broader research community. Furthermore, since they are specifically designed for problems that can be expressed in the QUBO format, their applicability is limited to optimization problems involving terms of at most quadratic order. 
Although it is possible to map higher-order terms into quadratic terms, the procedure requires additional steps involving the introduction of auxiliary variables and penalty constraints.\cite{babbushConstructionEnergyFunctions2014,choubisaAcceleratedChemicalSpace2023}
An alternative global optimizer that does not require specialized hardware and can directly handle higher-order terms is Tensor Train Optimizer (TTOpt).\cite{sozykinTTOptMaximumVolume2022,nikitinAreQuantumComputers2022} TTOpt is based on the Tensor Train (TT) decomposition, a technique that has gained widespread interest for compressing high-dimensional tensors and addressing the exponential growth in memory consumption and computational cost with the number of dimensions, which is commonly referred to as the curse of dimensionality.\cite{oseledetsTensorTrainDecomposition2011,zurekTensorTrainOptimization2025,caoAMUSETTICATensorBasedApproach2025,gelssQuantumDynamicsCoupled2025}
It has been demonstrated that TTOpt can solve complex QUBO problems involving thousands of variables and that it yields results comparable to those found with a digital annealer and other global optimizers.\cite{nikitinAreQuantumComputers2022}
 
In this work, we demonstrate that TTOpt is suitable for identifying optimal adsorption configurations and energies for practically relevant catalytic systems. To achieve this, we formulate the adsorption of simple molecules such as CO and NO on the surfaces of various binary alloy slabs as a higher-order binary optimization (HUBO) problem---an extension of the QUBO formulation that incorporates higher-order terms---and solve it using TTOpt. We further illustrate the versatility of our approach by applying it to high-entropy alloy (HEA) nanoparticles. These alloys are composed of five or more principal elements and have gained wide attention in recent years due to their high potential in catalysis, as well as other influential fields of chemistry and materials science.\cite{yehNanostructuredHighEntropyAlloys2004,cantorMicrostructuralDevelopmentEquiatomic2004,georgeHighentropyAlloys2019,wuPlatinumGroupMetalHighEntropyAlloyNanoparticles2020,duAlloyElectrocatalysts2023,ouyangRiseHighentropyBattery2024,shiotaLoweringExponentialWall2025, caiSurfaceengineeredNanostructuredHighentropy2025} 
Our results indicate that TTOpt is suitable for optimizing adsorption configurations without the drawbacks of quantum and digital annealers, such as the need for specialized hardware or the restriction to problems with at most quadratic terms. 
Furthermore, our work shows that higher-order terms are essential for accurately capturing multi-body interactions in catalytic systems, making the HUBO formulation in combination with TTOpt a promising approach for advancing catalyst design and other critical applications.

\section{Methods}

\begin{figure*}
    \includegraphics{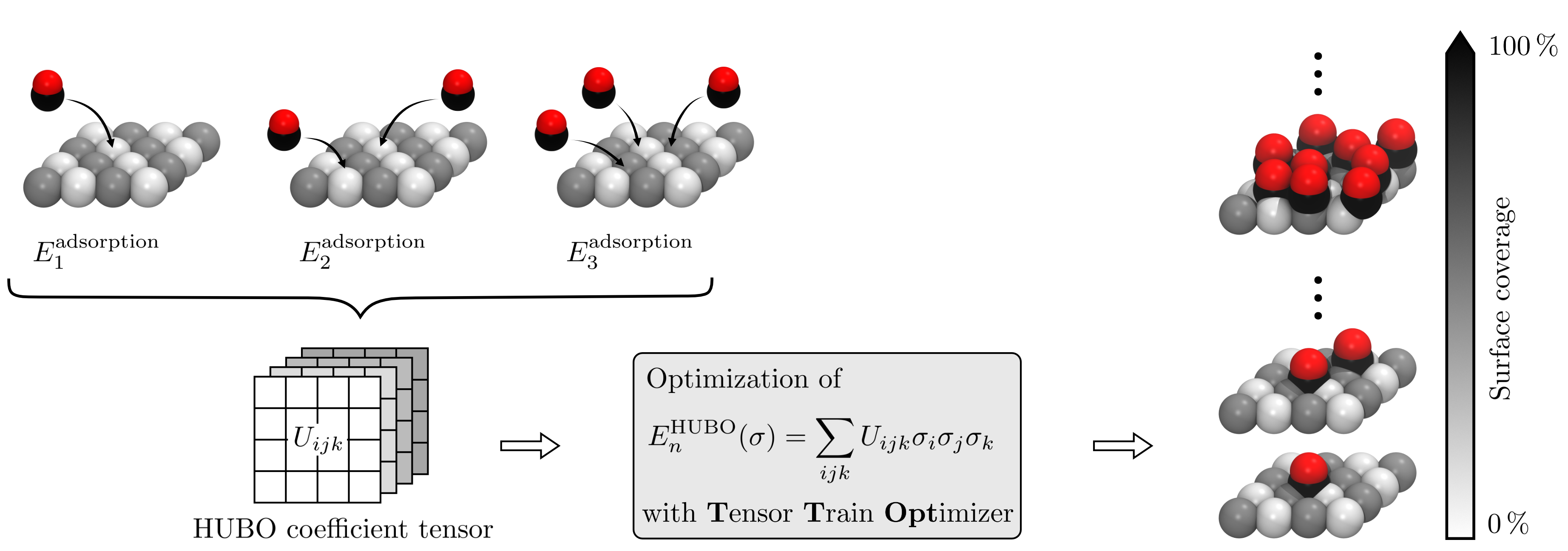}
    \centering
    \caption{\textbf{Schematic overview of the workflow.} The HUBO coefficient tensor was constructed by first computing all possible uni-, bi-, and trimolecular adsorption energies, which were then used in eqns~(\ref{eq:unimolecular_term_HUBO})–(\ref{eq:trimolecular_term_HUBO}) to obtain the coefficients $U_{ijk}$. The resulting HUBO formulation of the adsorption energy, eqn~(\ref{eq:adsorption_energy_3}), was then globally optimized using TTOpt to identify the most favorable adsorption configurations for surface coverages between 0 and \SI{100}{\%}.}
    \label{fig:Methods_overview}
\end{figure*}

\subsection{Theoretical framework}
For a substrate with $N$ adsorption sites, the adsorption configuration can be described by a vector $\sigma = (\sigma_1, \sigma_2, \dots, \sigma_N)$, where the occupation number $\sigma_i \in \{0,1\}$ for $i=1, \dots, N$ indicates whether adsorption site $i$ is unoccupied ($\sigma_i=0$) or occupied ($\sigma_i=1$) by an adsorbate.
The adsorption energy $E^\mathrm{adsorption}_n(\sigma)$ for a configuration $\sigma$ with $n$ adsorbates on the substrate surface is then given by
\begin{equation}
    E^\mathrm{adsorption}_n(\sigma) = E_n^\mathrm{total}(\sigma) - E^\mathrm{substrate} - n\cdot E^\mathrm{adsorbate} \,.
    \label{eq:adsorption_energy}
\end{equation}
$E^\mathrm{adsorbate}$ denotes the energy of a single adsorbate molecule in vacuum, $E^\mathrm{substrate}$ the energy of the substrate in vacuum, and $E_n^\mathrm{total}(\sigma)$ the energy of the substrate with $n$ adsorbates placed according to $\sigma$. 
An alternative expression for the adsorption energy can be obtained by representing it in terms of the occupation numbers of the adsorption sites:  
\begin{equation}
    E^\mathrm{adsorption}_n(\sigma) = \sum_{i_1=1}^{N}\sum_{i_2=1}^{N}\dots\sum_{i_n=1}^{N} U_{i_1 i_2 \dots i_n } \sigma_{i_1}\sigma_{i_2}\dots\sigma_{i_n} \,.
    \label{eq:hamiltonian}
\end{equation}
The coefficients $U_{i_1 i_2 \dots i_n}$ can be determined from the adsorption energies of 1 to $n$ adsorbates, obtained for all possible adsorption configurations. These coefficients are chosen such that the adsorption energies calculated using eqn~(\ref{eq:hamiltonian}) match those computed with eqn~(\ref{eq:adsorption_energy}). 

\subsubsection{QUBO formulation}
When the expansion in eqn~({\ref{eq:hamiltonian}}) is restricted to terms involving up to two adsorbates, the search for the global minimum configuration using the resulting expression is known as a quadratic unconstrained binary optimization (QUBO) problem. 
In this case, eqn~(\ref{eq:hamiltonian}) becomes\cite{sampeiQuantumAnnealingBoosts2023}
\begin{align}
    E^\mathrm{QUBO}_n(\sigma) & = \sum_{i=1}^N \sum_{j=1}^N U_{ij} \sigma_i \sigma_j \\
    & = \sum_{i=1}^N U_{ii}\sigma_i + 2\sum_{i=1}^N\sum_{j>i}^N U_{ij}\sigma_i\sigma_j \,.
    \label{eq:adsorption_energy_2}
\end{align}
In general, the adsorption energy obtained from eqn~(\ref{eq:adsorption_energy_2}) differs from that from eqn~(\ref{eq:hamiltonian}). Only for $n \leq 2$, $E^\mathrm{QUBO}_n(\sigma) = E^\mathrm{adsorption}_n(\sigma)$. The diagonal elements $U_{ii}$ and the off-diagonal elements $U_{ij}$ with $i\neq j$ of the QUBO coefficient matrix can be determined from unimolecular and bimolecular adsorption energies, respectively, as explained below.

When only one adsorbate is on the substrate and it occupies site $i=\alpha$, the resulting adsorption energy is given by
\begin{equation}
    E^\mathrm{QUBO}_1(\dots, \sigma_\alpha=1, \dots) = U_{\alpha\alpha} \,,
\end{equation}
where the notation $(\dots, \sigma_\alpha=1, \dots)$ indicates that all occupation numbers are zero except for site $i=\alpha$.
Therefore, the diagonal elements $U_{ii}$ of the QUBO coefficient matrix correspond to unimolecular adsorption energies $E^\mathrm{adsorption}_1$:
\begin{equation}
    U_{ii} = E^\mathrm{adsorption}_1(\dots, \sigma_i=1, \dots) \quad \text{for} \quad i = 1,\dots, N \,.
    \label{eq:unimolecular_term_1}
\end{equation}
When two adsorbates are present and they occupy sites $i=\alpha$ and $j=\beta$, eqn~(\ref{eq:adsorption_energy_2}) becomes
\begin{align}
    E^\mathrm{QUBO}_2(\dots, \sigma_\alpha=1, \dots, \sigma_\beta=1, \dots) &= \nonumber \\ 
    U_{\alpha\alpha} &+ U_{\beta\beta} + 2U_{\alpha\beta} \,.
    \label{eq:unimolecular_term_QUBO}
\end{align}
This shows that the off-diagonal elements $U_{ij}$ with $i \neq j$ can be determined from bimolecular adsorption energies $E^\mathrm{adsorption}_2$ according to
\begin{equation}
    \begin{aligned}
    U_{ij} = \frac{1}{2}(E^\mathrm{adsorption}_2(\dots, \sigma_i=1, \dots, \sigma_j=1, \dots) - U_{ii} - U_{jj}) \\
    \text{for} \quad i,j = 1,\dots, N \quad \text{and} \quad i \neq j \,,
    \label{eq:bimolecular_term_QUBO}
    \end{aligned}
\end{equation}
where $U_{ii}$ and $U_{jj}$ can be obtained from eqn~(\ref{eq:unimolecular_term_1}).
It has been demonstrated that the optimization of eqn~(\ref{eq:adsorption_energy_2}) can accurately predict the most stable adsorption configuration and the corresponding adsorption energy when the surface coverage is sufficiently low.\cite{sampeiQuantumAnnealingBoosts2023}
At high surface coverage, the adsorbates are close to each other, resulting in strong multi-body interactions that cannot be captured by the QUBO formulation in eqn~(\ref{eq:adsorption_energy_2}). To obtain more accurate results, it is necessary to include higher-order terms.

\subsubsection{HUBO formulation}
By incorporating terms involving three adsorbates, the QUBO formulation of the adsorption energy, eqn~(\ref{eq:adsorption_energy_2}), is extended to the following HUBO formulation:
\begin{align}
    E^\mathrm{HUBO}_n(\sigma) & = \sum_{ijk} U_{ijk} \sigma_i \sigma_j \sigma_k \\
    &= \sum_i U_{iii}\sigma_i + 6\sum_{i<j} U_{ijj}\sigma_i\sigma_j + 6\sum_{i<j<k} U_{ijk}\sigma_i\sigma_j\sigma_k \,.
    \label{eq:adsorption_energy_3}
\end{align}
This expression yields the exact adsorption energy for $n \leq 3$. 
The elements $U_{ijk}$ of the HUBO coefficient tensor can be obtained from unimolecular, bimolecular, and trimolecular adsorption energies, following the same procedure as for the elements $U_{ij}$ of the QUBO coefficient matrix discussed in the previous section.
The resulting expressions are:
\begin{align}
    U_{iii} &= E^\mathrm{adsorption}_1 \quad \text{for} \quad i = 1,\dots, N \label{eq:unimolecular_term_HUBO}\\
    U_{ijj} &= \frac{1}{6}(E^\mathrm{adsorption}_2 - U_{iii} - U_{jjj}) \nonumber \\
    &\text{for} \quad i,j = 1,\dots, N \quad \text{and} \quad i \neq j, \label{eq:bimolecular_term_HUBO}\\
    U_{ijk} &= \frac{1}{6} E^\mathrm{adsorption}_3 - U_{ijj} - U_{ikk} - U_{jkk} \nonumber\\
    &- \frac{1}{6}(U_{iii} + U_{jjj} + U_{kkk}) \nonumber \\
    &\text{for} \quad i,j,k = 1,\dots, N \quad \text{and} \quad i \neq j \neq k \,. \label{eq:trimolecular_term_HUBO}
\end{align}
For conciseness, we have omitted the explicit specification of the adsorption configuration $\sigma$.

Figure~\ref{fig:Methods_overview} provides a schematic overview of the workflow adopted in this work to identify optimal adsorption configurations using the HUBO formulation of the adsorption energy. First, all possible uni-, bi-, and trimolecular adsorption energies were computed and subsequently used to build the HUBO coefficient tensor according to eqns~(\ref{eq:unimolecular_term_HUBO})-(\ref{eq:trimolecular_term_HUBO}). Using the obtained coefficients $U_{ijk}$, the HUBO formulation of the adsorption energy, eqn~(\ref{eq:adsorption_energy_3}), was constructed. Finally, this objective function was optimized using TTOpt to identify the most favorable adsorption configurations across surface coverages from 0 to \SI{100}{\%}. To constrain the configuration to a fixed surface coverage, a penalty term was added to the cost function:
\begin{align}
    E^\mathrm{penalty} = E^\mathrm{penalty}_0(n^\mathrm{target}-n^\mathrm{trial})^2 \,.
\end{align}
$n^\mathrm{target}$ denotes the desired number of adsorbates and $n^\mathrm{trial}$ refers to the number of adsorbates in a given trial configuration. The prefactor $E^\mathrm{penalty}_0$ determines the steepness of the penalty and was set to \SI{10}{eV}.

For the QUBO formulation of the adsorption energy, eqn~(\ref{eq:adsorption_energy_2}), a similar procedure was followed. In this case, only uni- and bimolecular adsorption energies were determined, and eqns~(\ref{eq:unimolecular_term_1}) and (\ref{eq:bimolecular_term_QUBO}) were employed to construct the QUBO coefficent matrix.
The following section provides computational details for the determination of the adsorption energies and the optimization using TTOpt.

\subsection{Computational details}

\begin{figure}
    \includegraphics{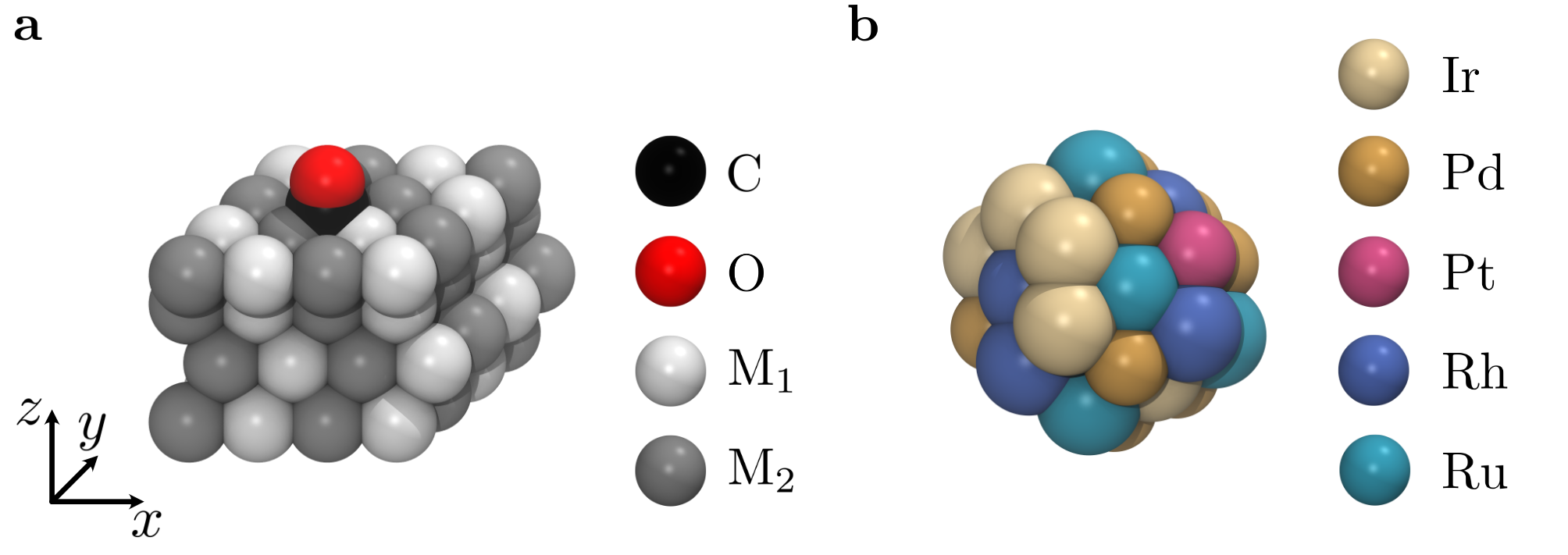}
    \centering
    \caption{\textbf{Illustration of the slab and nanoparticle structures.} (a) Structure of a binary (111) slab with a CO molecule adsorbed on its surface. The slab consists of $4\times 4 \times4$ atoms in the $x$-, $y$-, and $z$-directions, with a 1:1 ratio of elements M$_1$ and M$_2$. Periodic boundary conditions are applied in all directions, with a vacuum layer of $\SI{40}{\angstrom}$ along the $z$-direction. (b) Structure of one of the 25 IrPdPtRhRu HEA nanoparticles. The nanoparticle consists of 7 Ir, 8 Pd, 8 Pt, 7 Rh, and 8 Ru atoms randomly distributed within a truncated octahedral structure.
    }
    \label{fig:Adsorption_sites}
\end{figure}

\begin{figure*}
    \includegraphics{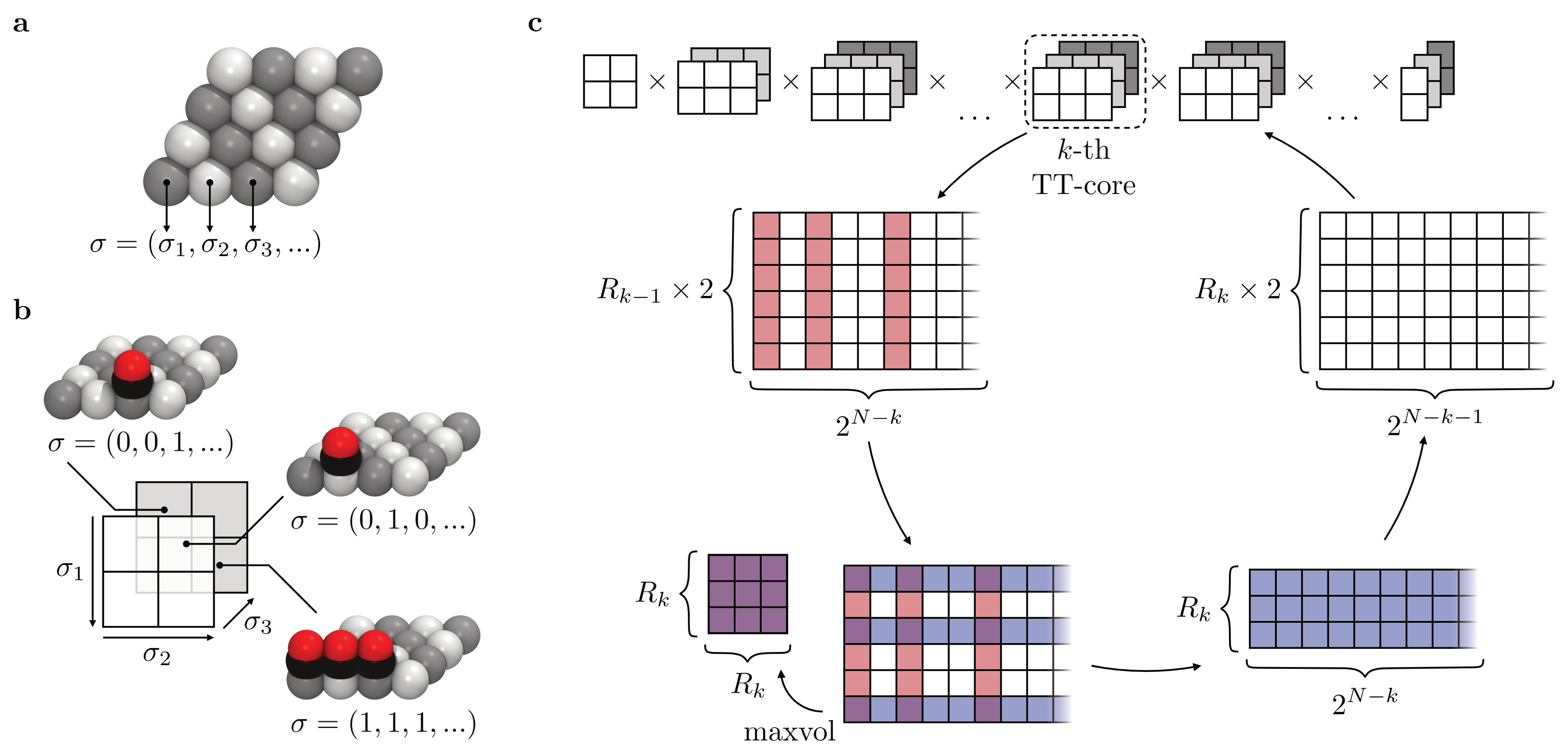}
    \centering
    \caption{\textbf{Optimization of the adsorption configurations with TTOpt.} (a) Mapping of the adsorption sites of a slab to the first three occupation numbers $\sigma_1$, $\sigma_2$, and $\sigma_3$ of the adsorption configuration $\sigma$. For simplicity, only the on-top sites are considered in this illustration. (b) Structure of the implicitly defined tensor $\mathcal{E}$ along $\sigma_1$, $\sigma_2$, and $\sigma_3$, with the corresponding adsorption configurations shown for three entries. The values of the entries are obtained from eqns~(\ref{eq:adsorption_energy_2}) and (\ref{eq:adsorption_energy_3}) for the QUBO and HUBO formulations, respectively. (c) Resulting TT with the maximal TT-rank set to 3. The $k$-th TT-core determines the possible occupation numbers at site $k$ by sampling multiple adsorption configurations conditioned on the choices made by the previous cores. The algorithm for selecting the possible occupation numbers is described in the main text.}
    \label{fig:TTOpt_illustration}
\end{figure*}

The energies $E_n^\mathrm{total}(\sigma)$, $E^\mathrm{substrate}$, and $E^\mathrm{adsorbate}$ in eqn~(\ref{eq:adsorption_energy}) were computed using the MACE neural network potential, an E(3)-equivariant graph neural network-based MLIP.\cite{batatiaMACEHigherOrder2022, batatiaDesignSpaceE3Equivariant2022} 
We employed MACE-Osaka24, a pre-trained universal MLIP capable of describing both organic molecular systems and inorganic crystal systems.\cite{shiotaTamingMultiDomainFidelity2024}
This model is available in three different versions -- \enquote{small}, \enquote{medium}, and \enquote{large} -- which differ in their parameter sizes.
For this study, we utilized the \enquote{large} version in double precision.

The initial geometries of the adsorbates, slabs, and nanoparticles were generated using the Atomic Simulation Environment (ASE).\cite{bahnObjectorientedScriptingInterface2002,larsenAtomicSimulationEnvironment2017} 
The initial adsorption configurations of $n$ adsorbates on the substrate surfaces  were constructed using the Alloy Catalysis Automated Toolkit (ACAT).\cite{hanRapidMappingAlloy2023} All geometry optimizations were performed with the limited memory version of the Broyden–Fletcher–Goldfarb–Shanno (LBFGS) algorithm as implemented in ASE and the threshold for the maximal force was set to \SI{e-3}{eV\per\text{\AA}}. The following sections provide a detailed description of the simulation procedure.

\subsubsection{Computation of the adsorption energy}
CO and NO were used as adsorbates. Their energies $E^\mathrm{adsorbate}$ were obtained from geometry-optimized structures in vacuum.
The substrates comprised slabs and nanoparticles.
The following binary alloys of the form M$_1$-M$_2$ with a 1:1 composition were used to model the slabs because of their relevance to applications involving CO and NO adsorption: 
Rh-Cu\cite{gonzalezTheoreticalStudyCO2005,liuRhCuAlloy2023,gonzalezTheoreticalStudyNO2007,geInterfacialEngineeringEnhances2022}, Rh-Ni\cite{leeDensityFunctionalTheory2010,zhongRhNiBimetallenesLatticeCompressed2024,wangInducingSynergyBimetallic2020}, Rh-Ru\cite{xiaoLeadEffectiveFacilitator2023,bagot3DAtomProbe2008}, Ir-Ni\cite{dagleSteamReformingHydrocarbons2016,wenMechanismNitricOxide2019}, Ir-Ru\cite{songPromotiveEffectsBa2023,baiInsightMechanismSelective2023}, Ni-Co\cite{liuFirstprinciplesStudyC+O2011,villagra-sozaScrutinizingMechanismCO22022,renNiCoAlloyParticledecorated2025}, Ni-Cu\cite{patelAtomicScaleSurfaceStructure2019,caiAmbientAmmoniaProduction2023}, Pd-Pt\cite{youPtPd111SurfacePtAu1112018,dollingProbingActiveSites2024,kimAlloyNanocatalystsElectrochemical2019,tangSinglePtPd2019,sarkerElectrocatalyticNitrateNitrous2022}, Pd-Au\cite{mazzoneInteractionCOPdAu1112008,serajPdAuAlloyNanoparticle2017}, Pd-Ag\cite{maFormationStabilityCO2009,linPdAgBimetallicElectrocatalyst2019,troutmanPdAgAlloyNanocatalysts2020}, Pd-Zn\cite{jeroroInteractionCOSurface2007,weiComparativeDensityFunctional2012,wangHydrogenintercalationPdZnBimetallene2025,weiComparativeDensityFunctional2012}, Pt-Ni\cite{muSynergeticEffectSurface2011,liangPtEnrichedPtNiAlloy2014,wuReactionNONiPt2011,kimAlloyNanocatalystsElectrochemical2019}, and Pt-Ru\cite{dimakisAttractionRepulsionMechanismCarbon2009,dimakisAdsorptionCarbonMonoxide2016,wangIncreasingElectrocatalyticNitrate2021}.
To construct the slabs, a face-centered cubic (fcc) structure containing $4\times 4 \times4$ atoms of the first element M$_1$ along the $x$-, $y$-, and $z$-directions was generated, with the (111) surface aligned parallel to the $x$-$y$-plane. Half of the atoms were then replaced with the second element M$_2$ to obtain the structure shown in Fig.~\ref{fig:Adsorption_sites}a. Periodic boundary conditions were applied in all directions to create the slab model, and a vacuum layer of $\SI{40}{\angstrom}$ was introduced along the $z$-direction to prevent interactions between periodic images. Geometry optimizations were carried out by adjusting the supercell dimensions while keeping the scaled atomic positions fixed, and the energy of the resulting structure was used as $E^\mathrm{substrate}$.
For the nanoparticles, IrPdPtRhRu HEA alloys were chosen for their promising catalytic properties.\cite{wuPlatinumGroupMetalHighEntropyAlloyNanoparticles2020, caiSurfaceengineeredNanostructuredHighentropy2025} To build the nanoparticles, a truncated octahedron consisting of 38 Pd atoms was first generated. 30 Pd atoms were then randomly replaced with 7 Ir, 8 Pt, 7 Rh, and 8 Ru atoms to approximate a 1:1:1:1:1 composition. This procedure was repeated to construct a total of 25 different nanoparticles. An example of a resulting nanoparticle is shown in Fig.~\ref{fig:Adsorption_sites}b. To preserve their structures, no geometry optimizations were performed. The energy of each resulting structure was used as $E^\mathrm{substrate}$. 

To obtain the energy of a slab with one adsorbate on its surface, an adsorbate molecule was placed at one of the possible adsorption sites of the previously generated slab using ACAT. All 96 adsorption sites on the (111) surface were determined using ACAT and are shown in Fig.~\ref{fig:Site_numbering_slab}.
A geometry optimization was performed with the positions of the slab atoms kept fixed, while the adsorbates were allowed to relax only perpendicular to the (111) surface. The energy of the optimized structure was used as $E_1^\mathrm{total}(\sigma)$. To compute the energy of slabs with two or three adsorbates, the corresponding number of adsorbates was placed on the selected adsorption sites using ACAT. The geometry optimization was conducted as described above to obtain $E_2^\mathrm{total}(\sigma)$ and $E_3^\mathrm{total}(\sigma)$, respectively. However, if the distance between any two adsorbates was less than \SI{2.3}{\angstrom}, if a configuration was equivalent to a previously evaluated one under permutation, or if the geometry optimization exceeded 100 steps, the evaluation of $E_n^\mathrm{total}(\sigma)$ was skipped, and $E^\mathrm{adsorption}_n(\sigma)$ was set to \SI{10}{eV}.
The positive sign reflects repulsion between the adsorbates, and the value was chosen to be much larger than the magnitude of the unimolecular adsorption energies, which are on the order of \SI{1}{eV}.

The corresponding uni-, bi-, and trimolecular adsorption energies $E^\mathrm{adsorption}_1$, $E^\mathrm{adsorption}_2$, and $E^\mathrm{adsorption}_3$, respectively, were determined according to eqn~(\ref{eq:adsorption_energy}).
This procedure was carried out for all possible configurations of $n=1,2,3$ adsorbates on the 96 adsorption sites of the (111) surface of the slabs to construct the QUBO coefficient matrices according to eqns~(\ref{eq:unimolecular_term_QUBO}) and (\ref{eq:bimolecular_term_QUBO}), and the HUBO coefficient tensors according to eqns~(\ref{eq:unimolecular_term_HUBO})–(\ref{eq:trimolecular_term_HUBO}). 

For the nanoparticle substrates, similar procedures were employed. The main differences compared to the slab models were the number of adsorption sites and the constraints used during geometry optimizations. In this case, a total of 170 adsorption sites on the nanoparticle surface were identified using ACAT and are shown in Fig.~\ref{fig:Site_numbering_NP}. During geometry optimizations, the positions of the nanoparticle atoms were kept fixed, and the adsorbates were allowed to relax only along the axis defined by the bond between the two atoms of the adsorbate.

To evaluate the error introduced by the truncation in the QUBO and HUBO formulations, the exact adsorption energy, as defined in eqn~(\ref{eq:adsorption_energy}), was calculated as follows. First, the optimal adsorption configurations $\sigma_\mathrm{opt}$ identified by TTOpt were reconstructed using ACAT. The resulting geometries were then optimized until the maximum force was smaller than \SI{e-3}{eV\per\angstrom}, while keeping the positions of the substrate atoms and the $x$- and $y$-coordinates of the adsorbates fixed. The energies of the optimized structures $E_n^\mathrm{total}(\sigma_\mathrm{opt})$ were used to obtain the adsorption energies $E^\mathrm{adsorption}_n(\sigma_\mathrm{opt})$ according to eqn~(\ref{eq:adsorption_energy}). The errors of the QUBO and HUBO formulations were then defined as $E^\mathrm{adsorption}_n(\sigma_\mathrm{opt}) - E^\mathrm{QUBO}_n(\sigma_\mathrm{opt})$ and $E^\mathrm{adsorption}_n(\sigma_\mathrm{opt}) - E^\mathrm{HUBO}_n(\sigma_\mathrm{opt})$, respectively.

\subsubsection{Optimization with TTOpt}
For a given $n$, the optimal adsorption sites were identified using TTOpt\cite{sozykinTTOptMaximumVolume2022, nikitinAreQuantumComputers2022}, an adaptive method based on the low-rank TT-decomposition\cite{oseledetsTensorTrainDecomposition2011}, the multidimensional cross approximation method in the TT-format (TT-CAM)\cite{oseledetsTTcrossApproximationMultidimensional2010}, and the maximum volume (maxvol) submatrix construction\cite{goreinovHowFindGood2010}. 
In TT format, the elements of a $d$-dimensional tensor $\mathcal{T}$ are expressed as:
\begin{equation}
\mathcal{T}_{i_1, i_2, \ldots, i_d} = 
\sum_{r_0=1}^{R_0} \sum_{r_1=1}^{R_1} \cdots \sum_{r_d=1}^{R_d}
\mathcal{C}^{(1)}_{r_0, i_1, r_1} \,
\mathcal{C}^{(2)}_{r_1, i_2, r_2} \,
\cdots \,
\mathcal{C}^{(d)}_{r_{d-1}, i_d, r_d},
\label{eq:tensor_train}
\end{equation}
that is, as a product of three-dimensional tensors $\mathcal{C}^{(k)}$ with $k=1,\dots,d$, referred to as TT-cores. The internal summation indices $r_k$ run from 1 to the corresponding TT-ranks $R_k$, which determine the efficiency and accuracy of the approximation. The TT decomposition is most beneficial when the TT-ranks are small. In this case, it is said that a low-rank approximation exists, and the TT representation is free from the curse of dimensionality.\cite{oseledetsTensorTrainDecomposition2011} 

TTOpt represents the QUBO and HUBO formulations of the adsorption energy as implicit $N$-dimensional tensors $\mathcal{E}$, where each dimension corresponds to one of the $N$ adsorption sites. Since each site can be either occupied or unoccupied, the size of each dimension is 2. The search for the configuration with the lowest adsorption energy is reformulated as identifying the tensor element with the lowest value. 
Fig.~\ref{fig:TTOpt_illustration} illustrates this with a simplified example, where only the on-top sites of a slab are considered as possible adsorption sites. Fig.~\ref{fig:TTOpt_illustration}a shows which adsorption sites correspond to the first three occupation numbers $\sigma_1$, $\sigma_2$, and $\sigma_3$ of the configuration $\sigma$. The structure of $\mathcal{E}$ along these three dimensions is depicted in Fig.~\ref{fig:TTOpt_illustration}b. Since each dimension has size 2, there are eight entries in total, each representing a unique adsorption configuration. For illustration, the configurations corresponding to three of the entries are explicitly shown. The values of the entries are computed using eqn~(\ref{eq:adsorption_energy_2}) for the QUBO formulation or eqn~(\ref{eq:adsorption_energy_3}) for the HUBO formulation. Note that the tensor is only implicitly defined, i.e., the tensor elements are not explicitly stored but are computed on demand. 

Based on the structure of $\mathcal{E}$ and the maximum TT-rank specified by the user, TTOpt constructs a tensor train as defined in eqn~(\ref{eq:tensor_train}). Fig.~\ref{fig:TTOpt_illustration}c shows the resulting structure when the maximum TT-rank is set to 3, i.e., $R_k \leq 3$.
The $k$-th TT-core decides which occupation numbers at site $k$ are passed to the next TT-core by evaluating the adsorption energies of multiple configurations, given the occupation numbers selected by the previous TT-cores. The algorithm for the selection is a modified version of TT-CAM and is schematically depicted in Fig.~\ref{fig:TTOpt_illustration}c for the $k$-th TT-core.
First, $R_k$ columns are explicitly evaluated from a matrix obtained by unfolding a subset of tensor elements from $\mathcal{E}$ along the $k$-th dimension, with the subset determined by the previous TT-cores.
Since each column contains $R_{k-1}\times2$ entries, a total of $R_k\times R_{k-1}\times2$ adsorption configurations are sampled.
The values in these columns are then transformed by a mapping function
\begin{equation}
    \mathcal{E}'(\sigma) = \frac{\pi}{2} - \arctan(\mathcal{E}(\sigma) - \mathcal{E}_\mathrm{min}),
    \label{eq:mapping_function}
\end{equation}
where the tensor element $\mathcal{E}(\sigma)$ is the adsorption energy of configuration $\sigma$, and $\mathcal{E}_\mathrm{min}$ denotes the current best estimate of the lowest adsorption energy. This transformation is necessary because the maxvol algorithm, which is used in the next step to identify large elements, does not consider the sign of the entries. The selected columns, after applying eqn~(\ref{eq:mapping_function}), are marked in red in the first step of the cycle. Using the maxvol algorithm, $R_k$ rows (blue) are selected to maximize the volume of the $R_k\times R_k$ submatrix formed by the intersection (purple) of the chosen $R_k$ rows and $R_k$ columns. In the next step, a matrix constructed from these rows is implicitly defined, then reshaped by unfolding along the next dimension and passed to the subsequent TT-core. These operations are repeated for the next TT-core, and once the final core is reached, the reshaping proceeds in the reverse direction, and adsorption configurations are sampled along the rows instead of the columns. 
This process continues until the number of times the adsorption energy is computed via eqn~(\ref{eq:adsorption_energy_2}) or~(\ref{eq:adsorption_energy_3}) reaches the user-specified maximum number of evaluations.

Since the TT-ranks determine how many rows and columns are selected at each step, the user-specified maximum TT-rank directly affects the sampling efficiency.
To optimize this parameter, one separate TTOpt run was performed for each maximum rank between 2 and 50, and only the best result was used for further analysis. The maximum number of evaluations for each run was set to \SI{e7}{}. The influence of the order of the site indices on the performance of TTOpt is discussed in the Appendix.

\section{Results}
\begin{figure*}
    \includegraphics{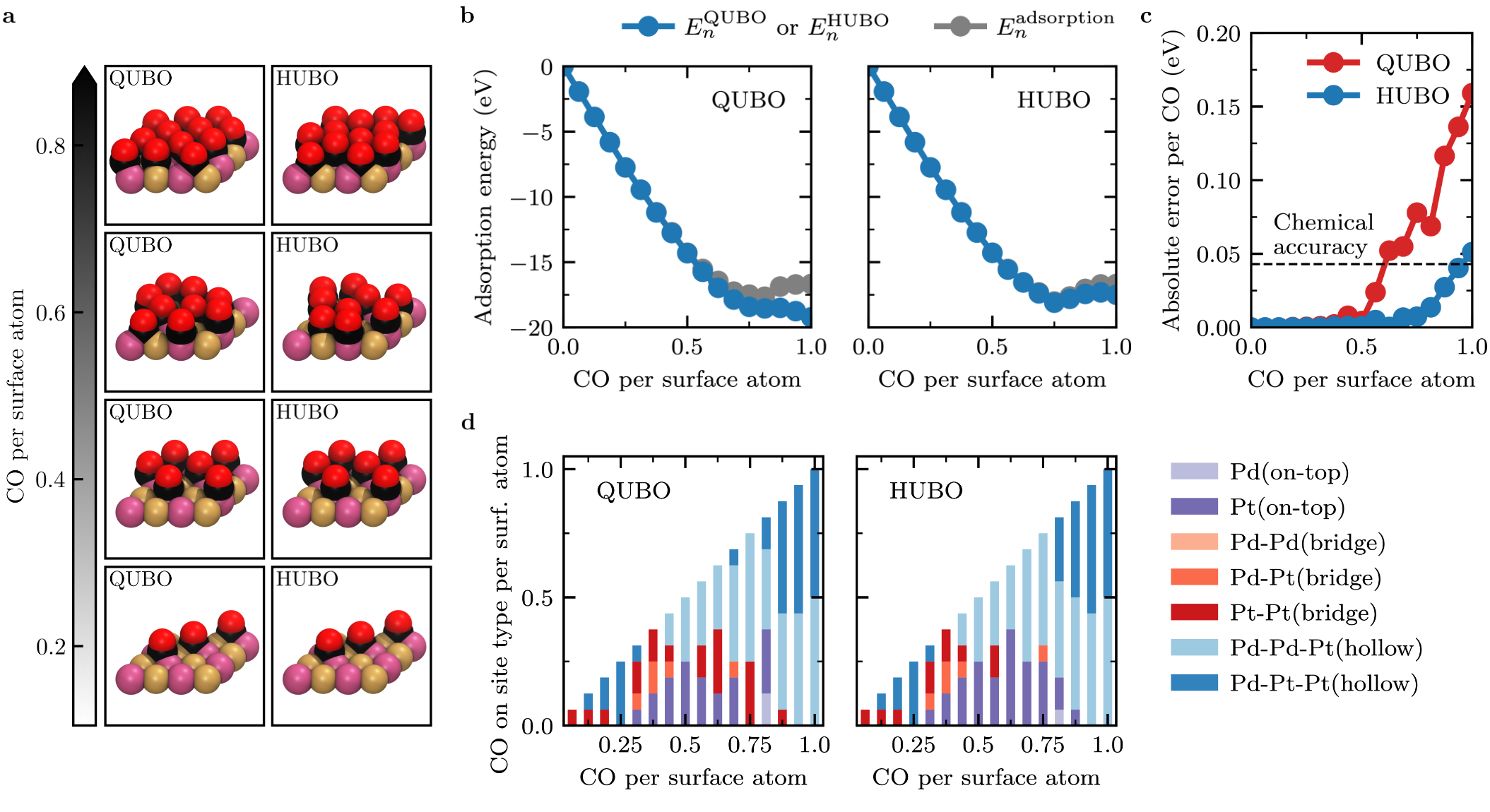}
    \centering
    \caption{\textbf{Comparison of optimal adsorption energies and configurations obtained from the QUBO and HUBO formulations for CO adsorption on a Pd-Pt slab.} (a) Illustration of the optimal adsorption configurations identified using the QUBO (left) and HUBO (right) formulations at 0.2, 0.4, 0.6, and 0.8 CO per surface atom. (b) Comparison of the QUBO and HUBO adsorption energies (blue) with the exact adsorption energies (grey). The deviation from the exact values increases with higher surface coverage and is more pronounced for the QUBO formulation.    
    (c) The inclusion of three-body interactions reduces the errors to below chemical accuracy ($\SI{0.043}{eV}$) even at surface coverages close to 1 CO per surface atom. The error was defined as the difference between the adsorption energy obtained from the QUBO or HUBO formulation and the exact expression for the same configuration. (d) The optimal adsorption configurations found using the QUBO formulation differ from those obtained with the HUBO formulation at surface coverages above 0.5 CO per surface atom. This indicates that the accuracies of the QUBO and HUBO formulations directly affects the configurations identified by TTOpt. Lines connecting the data points are drawn as guides for the eyes.}
    \label{fig:Adsorption_energies_QUBO_HUBO}
\end{figure*}

\subsection{CO adsorption on a Pd-Pt slab} 

To illustrate the effect of including three-body terms, we begin our analysis with the Pd-Pt slab as a representative example. 
Fig.~\ref{fig:Adsorption_energies_QUBO_HUBO}a presents the optimal adsorption configurations for 0.2, 0.4, 0.6, and 0.8 CO per surface atom obtained using the QUBO (left) and HUBO (right) formulations. For 0.2 and 0.4 CO per surface atom, the same configurations were identified, whereas at higher surface coverages, the adsorption configurations differ. This indicates that the three-body terms begin to significantly influence the adsorption energy only beyond 0.4 CO per surface atom.
To analyze this observation, we examine the adsorption energies, the errors introduced by the QUBO and HUBO formulations, as well as the distribution of the CO molecules on the Pd-Pt slab surface for surface coverages between 0 and 1 CO per surface atom.

Fig.~\ref{fig:Adsorption_energies_QUBO_HUBO}b shows the adsorption energies obtained from the QUBO (left) and HUBO (right) formulations, as determined by eqns~(\ref{eq:adsorption_energy_2}) and (\ref{eq:adsorption_energy_3}), respectively (blue curves).
The adsorption energies initially decrease for both formulations as the number of CO per surface atom increases. However, once the coverage exceeds 0.75 CO per surface atom, the QUBO and HUBO formulations predict different trends. For the QUBO formulation, the adsorption energy first remains nearly constant and then decreases with increasing surface coverage. In contrast, the adsorption energy increases for the HUBO formulation, indicating that the repulsive interactions between the CO molecules surpass the energy gained from additional CO adsorption, which is not captured by the QUBO formulation.

To validate the accuracy of our energy evaluations, we recalculated the adsorption energies of the optimal configurations identified by TTOpt for both the QUBO and HUBO formulations using eqn~(\ref{eq:adsorption_energy}) (grey curves). Note that the same configurations were used for the grey and blue curves within each formulation, while the optimal configurations identified by TTOpt for QUBO and HUBO may differ. For the QUBO formulation, the approximated adsorption energies start to deviate from the exact expression once the coverage exceeds 0.5 CO per surface atom, with the magnitude of deviation increasing further at higher coverages. At surface coverages above 0.75 CO per surface atom, opposite trends are observed, with the QUBO adsorption energy decreasing, while the exact energies increase. In contrast, the deviations of the adsorption energies are significantly smaller for the HUBO formulation across the whole surface coverage range. Furthermore, the HUBO formulation reproduces the trend of the exact expression, demonstrating that the inclusion of third-order terms is essential for accurately capturing the interactions between CO molecules and reflecting site-saturation effects. 

The absolute errors per CO molecule for the QUBO (red) and HUBO (blue) formulations are quantified in Fig.~\ref{fig:Adsorption_energies_QUBO_HUBO}c, with the black dashed line representing the chemical accuracy threshold of $\SI{0.043}{eV}$. The error is defined as the difference between the adsorption energies of the QUBO and HUBO formulations and the respective adsorption energies obtained from MACE-Osaka24. The QUBO formulation maintains a low error for surface coverages below 0.5 CO per surface atom. However, at higher surface coverages, the absolute error increases sharply and exceeds the chemical accuracy threshold when the coverage is 0.6 CO per surface atom or higher. In contrast, the error of the HUBO formulation remains below the chemical accuracy threshold, and only starts to increase notably once the surface coverage exceeds saturation.

Fig.~\ref{fig:Adsorption_energies_QUBO_HUBO}d shows the site occupancy of CO per surface atom for the on-top, bridge, and hollow adsorption sites. These site types are illustrated in Fig.~\ref{fig:Site_numbering_slab}. For coverages below 0.6 CO per surface atom, the distribution of the CO molecules across the different adsorption site types is identical for the QUBO (left) and HUBO (right) formulations. However, at higher surface coverages, the distributions of CO molecules differ between the QUBO and HUBO formulations, demonstrating that TTOpt identifies different optimal adsorption configurations depending on the accuracy of the energy expression used as the cost function.

Since optimizing an inaccurate expression can result in less favorable adsorption configurations, it is important to include higher-order terms to accurately approximate the exact adsorption energy. This is illustrated by the following example: at a surface coverage of 0.75 CO per surface atom, the adsorption energy obtained from the QUBO formulation ($\SI{-18.422}{eV}$) is lower than that from the HUBO formulation ($\SI{-18.104}{eV}$). However, recalculations with MACE-Osaka24 show that the optimal configuration identified using the HUBO formulation yields a lower adsorption energy ($\SI{-18.016}{eV}$) than that identified using the QUBO formulation ($\SI{-17.487}{eV}$).

\begin{figure*}
    \includegraphics{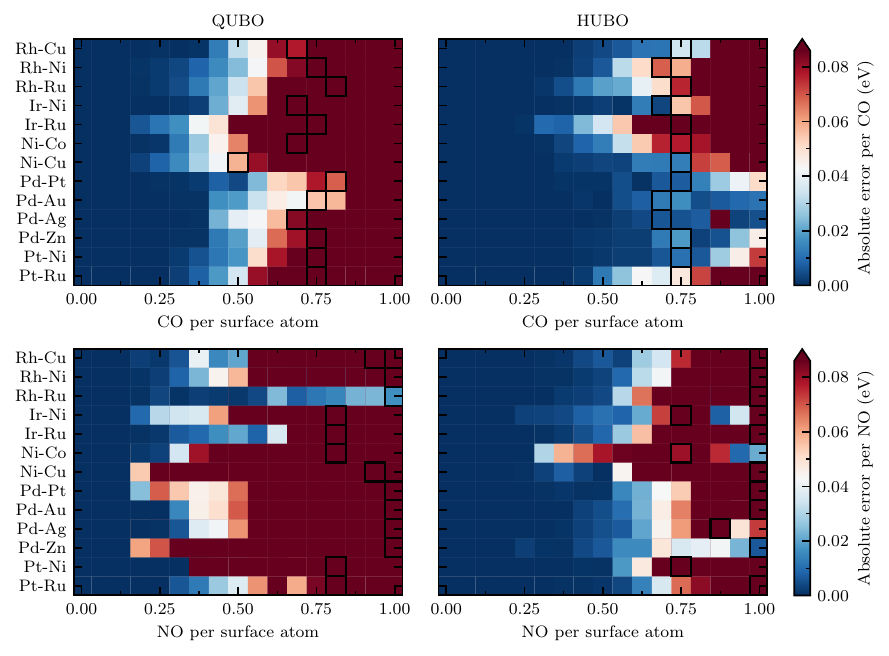}
    \centering
    \caption{\textbf{Adsorption energy errors of the optimal configurations identified with the QUBO and HUBO formulations.} The absolute error per CO (top) and NO (bottom) resulting from the QUBO (left) and HUBO (right) formulations was determined for various binary alloys. The black squares indicate the surface coverage at which the lowest adsorption energy was found across the whole coverage range. In general, the inclusion of third-order terms extends the surface coverage range over which the error per adsorbate molecule remains below the chemical accuracy threshold of \SI{0.043}{eV}.}
    \label{fig:QUBO_HUBO_errors}
\end{figure*}

\begin{figure*}
    \includegraphics{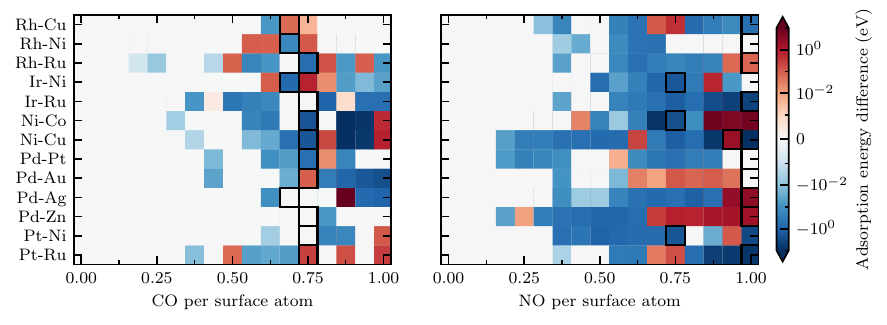}
    \centering
    \caption{\textbf{Adsorption energy differences of the optimal configurations identified with the QUBO and HUBO formulations.} The differences in the CO (left) and NO (right) adsorption energies of the optimal configurations identified by TTOpt using the QUBO and HUBO formulations as cost functions was determined for various binary alloys. The black squares mark the surface coverage at which the lowest adsorption energy was found across the whole coverage range. If this value differs between the QUBO and HUBO results, the lower-energy configuration was used. The results reveal that the HUBO formulation generally leads to more favorable adsorption configurations. This shows that incorporating higher-order terms is important not only for improving accuracy but also for identifying optimal adsorption configurations.}
    \label{fig:QUBO_HUBO_differences}
\end{figure*}

\subsection{Extension to NO adsorption and additional binary alloys}
We extended our simulations to NO as the adsorbate and to additional binary alloys to demonstrate the general validity of our findings.
For simplicity and to ensure comparability, we neglected surface reconstruction and assumed that the adsorbates align perpendicular to the slab surface. Fig.~\ref{fig:QUBO_HUBO_errors} presents the absolute error per CO (top) and NO (bottom) for all studied binary alloys for surface coverages between 0 and 1 adsorbates per surface atom. 
The black squares mark the surface coverage of the configuration with the lowest adsorption energy over the whole coverage range. The energies were evaluated using the exact expression defined in eqn~(\ref{eq:adsorption_energy}) after optimization with TTOpt using the QUBO (left) or HUBO (right) formulation.

The error is compared to the chemical accuracy threshold of $\SI{0.043}{eV}$, with blue indicating values below this threshold, while white represents values equal to it and red those exceeding it. We begin our analysis with CO as the adsorbate. For the QUBO formulation, the error is below the chemical accuracy threshold up to approximately 0.4 to 0.6 CO per surface atom, depending on the alloy. 
When three-body interactions are included in the HUBO formulation, this range is extended in all cases. For more than half of the studied alloys, the error remains within chemical accuracy even at the surface coverage corresponding to the configuration with the lowest energy (black squares), i.e., when the surfaces are saturated.
This demonstrates the importance of three-body terms for accurately capturing the interactions between CO molecules and suggests that they may be sufficient for modeling adsorption up to one monolayer coverage.

To illustrate the impact of including three-body terms on identifying optimal adsorption configurations, Fig.~\ref{fig:QUBO_HUBO_differences} shows the differences in adsorption energies between the configurations identified by TTOpt using the QUBO and HUBO formulations as cost functions. The energies were evaluated with the exact expression given in eqn~(\ref{eq:adsorption_energy}). The color scheme reflects which formulation results in lower adsorption energies: blue for HUBO, red for QUBO, and white indicates that both formulations lead to optimal configurations with the same adsorption energy. The surface coverage at which the configuration with the lowest adsorption energy over the whole coverage range was found is indicated by the black squares. If this coverage differs between the QUBO and HUBO formulations, the one corresponding to the configuration with the lower energy was chosen.

Fig.~\ref{fig:QUBO_HUBO_differences} (left) presents the results for CO as the adsorbate. For surface coverages up to 0.3 CO per surface atom, the adsorption energies of the optimal configurations obtained from both formulations are identical for all alloys except Rh-Ru, where the HUBO formulation yields lower-energy structures around 0.2 CO per surface atom. With increasing surface coverage, configurations with lower adsorption energies can be found for more alloys when the HUBO formulation is used as the cost function. This is most prevalent around 0.6 CO per surface atom, where the HUBO formulation results in configurations with lower adsorption energies for more than half of the alloys. 
Although there are cases where the QUBO formulation leads to adsorption configurations with lower adsorption energies, especially at surface coverages above saturation (black squares), this occurs in less than \SI{25}{\%} of the alloys at any coverage below saturation.

For NO as the adsorbate, similar trends are observed. The error per NO (Fig.~\ref{fig:QUBO_HUBO_errors}, bottom) in the QUBO formulation is below the chemical accuracy threshold between 0.2 and 0.5 NO per surface atom, while this range extends to at least 0.6 NO per surface atom in the HUBO formulation for most alloys. The differences in adsorption energies (Fig.~\ref{fig:QUBO_HUBO_differences}, right) reveal that more favorable adsorption configurations are obtained with the HUBO formulation for all alloys across nearly the entire surface coverage range. Although there are exceptions, most notably Pd-Zn and Pd-Au, these generally occur at surface coverages close to or above saturation (black squares).

In summary, this benchmark across multiple adsorbates and binary alloy surfaces consistently demonstrates the importance of extending the QUBO formulation to include higher-order terms to accurately approximate adsorption energies and identify optimal adsorption configurations using TTOpt.

\begin{figure*}
    \includegraphics{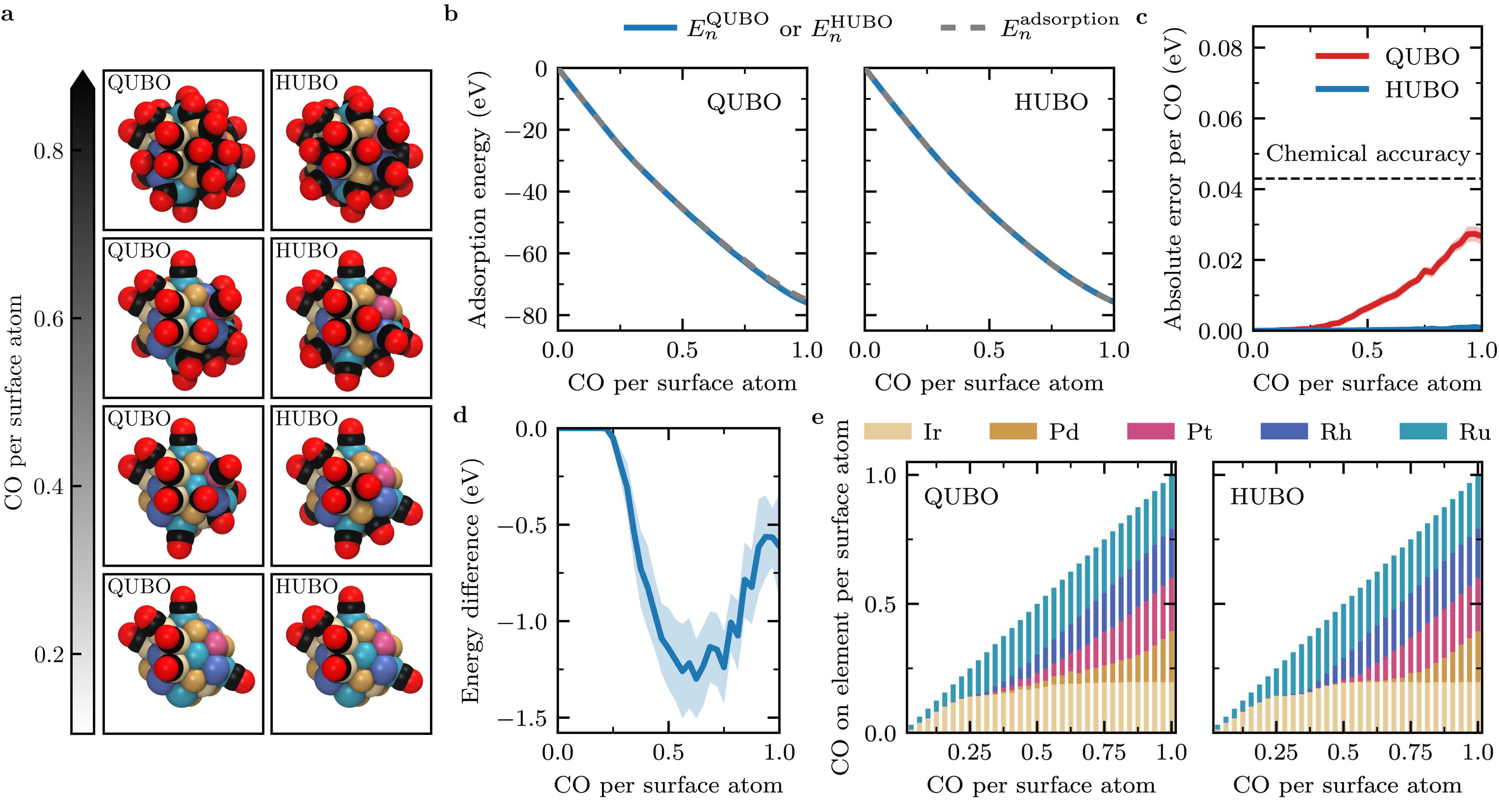}
    \centering
    \caption{\textbf{Comparison of optimal adsorption energies and configurations obtained from the QUBO and HUBO formulations for CO adsorption on an HEA nanoparticle.} (a) The studied nanoparticle is a truncated octahedron composed of 38 atoms, with Ir, Pd, Pt, Rh, and Ru randomly assigned while maintaining an equiatomic composition. (b) The adsorption energies obtained from both the QUBO (left) and HUBO (right) formulations decrease monotonically across the surface coverage range from 0 to 1 CO per surface atom. 
    The deviation between the QUBO and HUBO adsorption energies (blue) and the respective MACE-Osaka24 energies (grey) remains small even for high surface coverage. (c) The inclusion of three-body interactions decreases the errors close to zero. (d) The HUBO formulation yields more favorable adsorption configurations than the QUBO formulation at surface coverages above 0.25 CO per surface atom, as indicated by the adsorption energy differences of the respective optimal configurations. (e) 
    At surface coverages above 0.25 CO per surface atom, the optimal adsorption configurations identified using the QUBO formulation differ from those obtained with the HUBO formulation. This demonstrates the importance of including third-order terms for finding more favorable adsorption configurations with TTOpt. The shaded areas represent twice the standard error (\SI{95}{\%} confidence interval).}
    \label{fig:NP_results}
\end{figure*}

\subsection{Application to HEA nanoparticles}

Finally, we show that our method is not limited to ideal flat surfaces but is also applicable to highly complex systems. As a representative example, we chose HEA nanoparticles -- a class of advanced materials that have received growing interest in recent years for their unique properties, making them promising candidates for a broad range of applications. Since they are composed of five or more different elements, the resulting nanoparticles have complex, low-symmetry configurations with a high number of non-equivalent adsorption sites, making them an ideal benchmark for our method. 

In Fig.~\ref{fig:NP_results}a, we show the optimal adsorption configurations obtained with TTOpt using the QUBO (left) and HUBO (right) formulations for 0.2, 0.4, 0.6, and 0.8 CO per surface atom. While the QUBO and HUBO formulations yield the same adsorption configuration at 0.2 CO per surface atom, they lead to different configurations at higher surface coverages. This suggests that three-body interactions already become relevant once the coverage exceeds 0.2 CO per surface atom.
In the following, we compare how the extension of the QUBO formulation to the HUBO formulation affects the adsorption energies, deviations from the exact expression, quality of the identified configurations, and distribution of the CO molecules on the HEA nanoparticle surface for surface coverages between 0 and 1 CO per surface atom.

The adsorption energies of the optimal adsorption configurations obtained with TTOpt using the QUBO (left) and HUBO (right) formulations are shown in Fig.~\ref{fig:NP_results}b for surface coverages between 0 and 1 CO per surface atom. The blue curves depict the adsorption energies determined by eqns~(\ref{eq:adsorption_energy_2}) and (\ref{eq:adsorption_energy_3}). For both formulations, the adsorption energies decrease monotonically as the number of CO per surface atom increases. As the surface coverage approaches 0.5 CO per surface atom, the change in adsorption energy becomes smaller, which may result from saturation of the most favorable adsorption sites and increasing repulsion between adsorbed CO molecules. The grey curves represent the adsorption energies recalculated using the exact expression eqn~(\ref{eq:adsorption_energy}) for the optimal configurations identified by TTOpt and show good agreement with the approximated adsorption energies obtained from the QUBO and HUBO formulations. 
 
The absolute errors per CO molecule in the adsorption energies resulting from the QUBO and HUBO formulations are plotted in Fig.~\ref{fig:NP_results}c, with the chemical accuracy threshold of $\SI{0.043}{eV}$ indicated by the horizontal black horizontal dashed line. The error in the QUBO formulation (red) increases gradually with higher surface coverage but remains below the chemical accuracy threshold, even at a coverage of 1 CO per surface atom. However, the HUBO formulation maintains errors significantly below the chemical accuracy threshold and consistently close to zero throughout the entire CO coverage range. Compared to the slab models discussed in the previous sections, the errors for the nanoparticle are notably smaller, especially for the HUBO formulation. This result suggests that our method performs better for certain geometries that deviate from ideal flat surfaces, as the non-parallel orientation of CO molecules can increase the distance and therefore reduce the repulsion between them.

In Fig.~\ref{fig:NP_results}d, we show the difference in adsorption energy between the optimal adsorption configurations found with the HUBO and QUBO formulations, with the energies recalculated using the exact expression given in eqn~(\ref{eq:adsorption_energy}). At low surface coverages, no difference is observed. However, once the surface coverage exceeds 0.25 CO per surface atom, TTOpt consistently identifies adsorption configurations with lower energies when the HUBO formulation is used as the cost function. These findings demonstrate the importance of including third-order terms, not only for accurately capturing interactions between CO molecules and surface atoms but also for identifying optimal adsorption configurations when applying our method to highly complex systems.

Lastly, we discuss the distribution of the CO molecules on the surface of the HEA nanoparticle in Fig.~\ref{fig:NP_results}e. The number of CO molecules adsorbed per surface element is identical for the QUBO and HUBO formulations up to a surface coverage of 0.25 CO per surface atom. For both formulations, the CO molecules adsorb exclusively on Ru and Ir atoms within this coverage range, indicating a competitive preference between these two elements. When the surface coverage exceeds 0.25 CO per surface atom, the sites containing Rh, Pt, and Pd atoms also begin to compete for CO adsorption in the QUBO formulation. In contrast, for the HUBO formulation, the CO molecules first occupy all adsorption sites formed solely by Ir and Rh atoms before adsorbing onto sites that include Ru atoms. Once these are saturated, the CO molecules begin to occupy sites containing Pt atoms, which then compete with sites containing Pd atoms for CO adsorption. This behavior is consistent with the observed trend in adsorption energies, which decrease from Ir and Ru to Rh, and are lowest for Pd and Pt.\cite{shiotaLoweringExponentialWall2025}

\begin{figure}
    \includegraphics{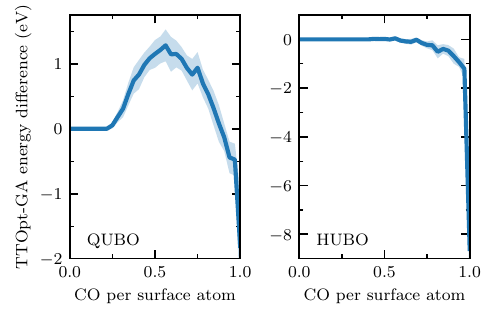}
    \centering
    \caption{\textbf{Difference in adsorption energy between the configurations identified by TTOpt and GA.} Negative values indicate that TTOpt identified more stable configurations, while positive values indicate that GA performed better. For the QUBO formulation (left), GA outperforms TTOpt at moderate coverages, whereas TTOpt performs better at high coverages. For the HUBO formulation (right), TTOpt consistently identifies lower-energy configurations across all coverages. The shaded areas represent twice the standard error (\SI{95}{\%} confidence interval).}
    \label{fig:TTOpt_GA}
\end{figure}

\subsection{Comparison with GA}
Before concluding this work, we compare TTOpt against genetic algorithm (GA), which is a well-established global optimization method that mimics natural evolution to optimize a target function.\cite{hollandGeneticAlgorithms1992} GA operates on a population of candidate solutions, selects individuals based on their fitness, combines pairs of solutions to produce offspring, introduces mutation with a certain probability, and repeats this process over multiple generations to explore the solution space. As such, GA involves several hyperparameters, including population size, mutation rate, and offspring size, which must be adjusted for optimal performance.

We repeated the search for the optimal adsorption configurations of the HEA nanoparticle using GA, implemented with the DEAP framework.\cite{fortinDEAPEvolutionaryAlgorithms2012} The population size was varied between 100, 300, 500, and 800, while an offspring size of 500, 1000, 1500, 2000, or 2500 was chosen. The mutation probability was set to 0.1, 0.3, 0.5, 0.7, or 0.9. The number of generations was adjusted so that the total number of evaluations reached \SI{e7}{}. This resulted in 76 GA searches for each HEA nanoparticle and each given number of adsorbed CO molecules, and only the best solution was used for further analysis.

Fig.~\ref{fig:TTOpt_GA} shows the difference in adsorption energy between the configurations identified by the two methods, with negative values and positive values indicating more stable configurations from TTOpt and GA, respectively. Note that since we are evaluating the performance of the two methods in finding the global minimum, the truncated expressions given in eqns~(\ref{eq:adsorption_energy_2}) and (\ref{eq:adsorption_energy_3}) were used here for the comparison. For the QUBO formulation (left), GA identifies the same adsorption configurations as TTOpt when the surface coverage is below 0.2 CO per surface atom. However, at surface coverages between 0.2 and 0.9 CO per surface atom, GA identifies energetically more stable configurations, whereas at higher surface coverages, TTOpt outperforms GA. For the HUBO formulation (right), the same adsorption configurations are identified up to approximately 0.5 CO per surface atom. At higher surface coverages, TTOpt consistently yields more stable configurations. 

These results show that TTOpt consistently outperforms GA at high surface coverages, where strong repulsive interactions between CO molecules considerably reduce the number of stable configurations. Since TTOpt is based on TT decomposition, it can learn the correlations between sites, allowing it to filter out unstable configurations early and focus the optimization on the energetically stable ones. 

However, at moderate surface coverages around 0.5 CO per surface atom, GA consistently identifies configurations with lower energy when using the QUBO formulation. Given that this does not occur with the HUBO formulation, this observation might stem from the larger errors introduced by the QUBO formulation, which are quantified in Fig.~\ref{fig:NP_results}c and discussed in the previous section.
The QUBO formulation distorts the energy landscape by lowering and increasing the energy of certain structures, potentially introducing artificially deep minima and high energy barriers. Our findings indicate that GA, being a stochastic method, can escape such artificial minima more easily than TTOpt, and therefore outperforms TTOpt in these cases. 

In contrast, TTOpt outperforms GA when the HUBO formulation is used. This suggests that the inclusion of third-order terms leads to a smoother and more faithful approximation of the true potential energy surface. Consequently, the HUBO formulation better preserves correlations between the adsorption sites than the QUBO formulation. Our results indicate that TTOpt can then exploit these correlations to navigate the energy landscape more effectively during optimization and reliably identify configurations with lower energy.

\section{Discussion}
In this work, we demonstrated that TTOpt is suitable for optimizing adsorption configurations when the adsorption energies are formulated as a HUBO problem. We observed that expressing the adsorption energy using only first- and second-order terms could limit the applicability to low surface coverages, and in some cases, the error already exceeds the chemical accuracy threshold of $\SI{0.043}{eV}$ for surface coverages of 0.2 adsorbates per surface atom. Incorporating third-order interaction terms greatly reduces errors, expanding the surface coverage range within chemical accuracy to between 0.5 and 1 adsorbates per surface atom for most of the studied alloys.

However, it should be noted that the number of HUBO coefficients grows exponentially with the order of the interaction terms. As such, the construction of the cost function itself is the most time-consuming step. For example, the construction of the QUBO matrix for one HEA nanoparticle took approximately \SI{3}{h}, whereas building the HUBO tensor required \SI{250}{h} of walltime. Both computations were performed on an NVIDIA H100 GPU configured with Multi-Instance GPU (MIG), utilizing a single instance, which corresponds to 1/7th of the total GPU capacity. The subsequent optimizations with TTOpt were carried out on an Intel Xeon Platinum 8490H processor. A search with the QUBO formulation as the cost function was performed on a single core and took \SI{3}{min}. For the HUBO formulation, the optimization was parallelized across 10 cores and completed in \SI{50}{min} of walltime, corresponding to a total CPU time of \SI{500}{min}.

Nevertheless, our results suggest that including up to third-order terms is sufficient for accurately capturing multi-body interactions in most application-relevant cases. Most notably, our benchmarks revealed that the errors in adsorption energy are reduced to near zero for nanoparticles once third-order terms are incorporated. Given that third-order terms only affect the identified adsorption configurations at surface coverages of 0.2 adsorbates per surface atom and higher, i.e., when the adsorbates are sufficiently close, introducing cutoff schemes could significantly reduce the computational cost of constructing the HUBO tensor. For instance, the minimum distance between the CO molecules on the Pd-Pt surface is \SI{3.7}{\angstrom} and \SI{2.8}{\angstrom} at coverages of 0.5 and 0.75 CO per surface atom, respectively, suggesting that a cutoff radius between these values might be sufficient for this system.

\section{Conclusion}
Our extensive benchmarks across various adsorbates, alloys, and surface geometries demonstrate that TTOpt can solve HUBO problems for chemical systems in a reasonable amount of time. Furthermore, our results highlight the importance of higher-order terms for accurately approximating adsorption energies and identifying optimal configurations using TTOpt. Unlike quantum and digital annealers, which require reformulations of the cost function to handle third- and higher-order terms, TTOpt can incorporate them directly and does not require specialized hardware. Instead, it can be readily used on widely accessible computational resources, including regular consumer desktops and laptops. Comparisons with GA, a well-established global optimization algorithm, revealed that GA can outperform TTOpt in certain cases when the QUBO formulation is used. Since this was not observed with the HUBO formulation, our results suggest that third-order terms are essential for preserving site correlations, which enhances the performance of TTOpt.

The scope of the present work is limited to HUBO formulations that include interaction terms up to third order. Extending the formulation to fourth- and higher-order terms remains challenging due to the computational cost, even when MLIPs are employed. To construct HUBO problems with such higher-order interactions more efficiently, approximate concepts may be included. One such approach is the local surface energy (LSE) descriptor introduced by Shiota and Ishihara et al.,\cite{shiotaLoweringExponentialWall2025} which estimates adsorption energies based on atomic energies obtained from the final layer of MLIPs. With LSE, the third-order terms could be predicted from pairwise interactions, which would reduce the cost of constructing the HUBO tensor to that of the QUBO matrix. Another promising strategy is the use of (higher-order) factorization machines to approximate the HUBO tensor as a product of vectors, which allows many-body interactions to be captured without computing every element of the tensor explicitly.\cite{blondelHigherOrderFactorizationMachines2016,kitaiDesigningMetamaterialsQuantum2020} These approximations could extend the applicability of our approach to larger systems and enable its use as a low-cost screening method for technological applications in catalysis, sensing, and related fields.

Our findings highlight the practical value of QUBO and HUBO formulations for adsorption problems in chemistry.
We anticipate that our work will contribute to the development of novel and practical methods in both classical and quantum computing.

\section{Acknowledgments}
We are grateful to Yuichiro Yoshida and Kenji Ishihara of the University of Osaka for fruitful discussions.
This work was supported by MEXT Quantum Leap Flagship Program (MEXTQLEAP) Grant No. JPMXS0120319794 and the JST COI-NEXT Program Grant No. JPMJPF2014. This research was partially facilitated by the JSPS Grants-in-Aid for Scientific Research (KAKENHI) Grant No. JP23H03819.
Calculations were performed using the Genkai supercomputer of the Research Institute for Information Technology at Kyushu University and the SQUID supercomputer at the Cybermedia Center of the University of Osaka.

\setcounter{equation}{0}
\setcounter{figure}{0}
\renewcommand{\theequation}{A\arabic{equation}}
\renewcommand{\thefigure}{A\arabic{figure}}

\section{Appendix}

\begin{figure}
    \includegraphics{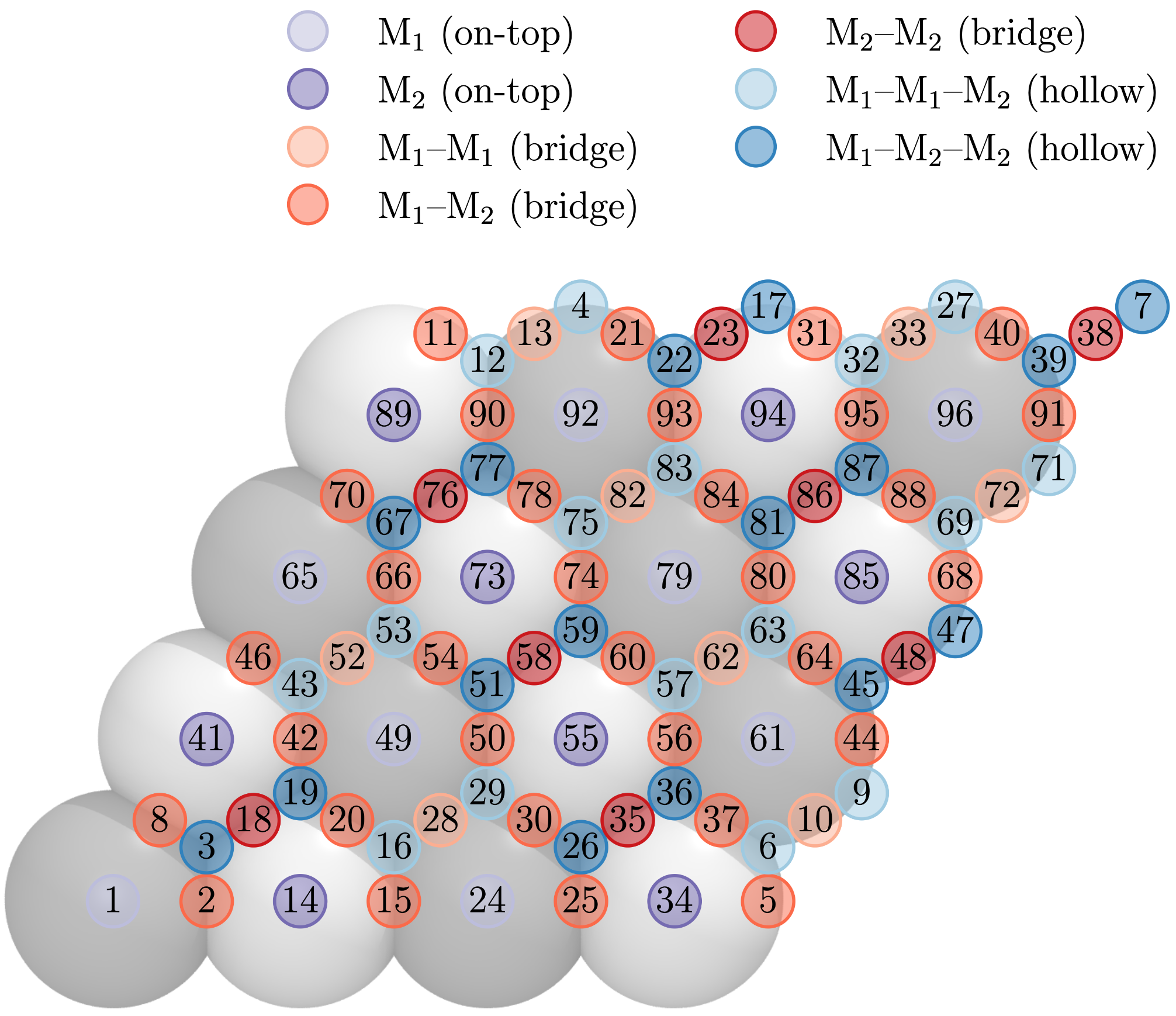}
    \centering
    \caption{\textbf{Enumeration of adsorption sites on the slab model.} All adsorption sites were identified and enumerated using ACAT. Each site is classified according to the number of surface atoms involved in the adsorption process: on-top (one atom), bridge (two atoms), and hollow (three atoms).}
    \label{fig:Site_numbering_slab}
\end{figure}

\begin{figure*}
    \includegraphics{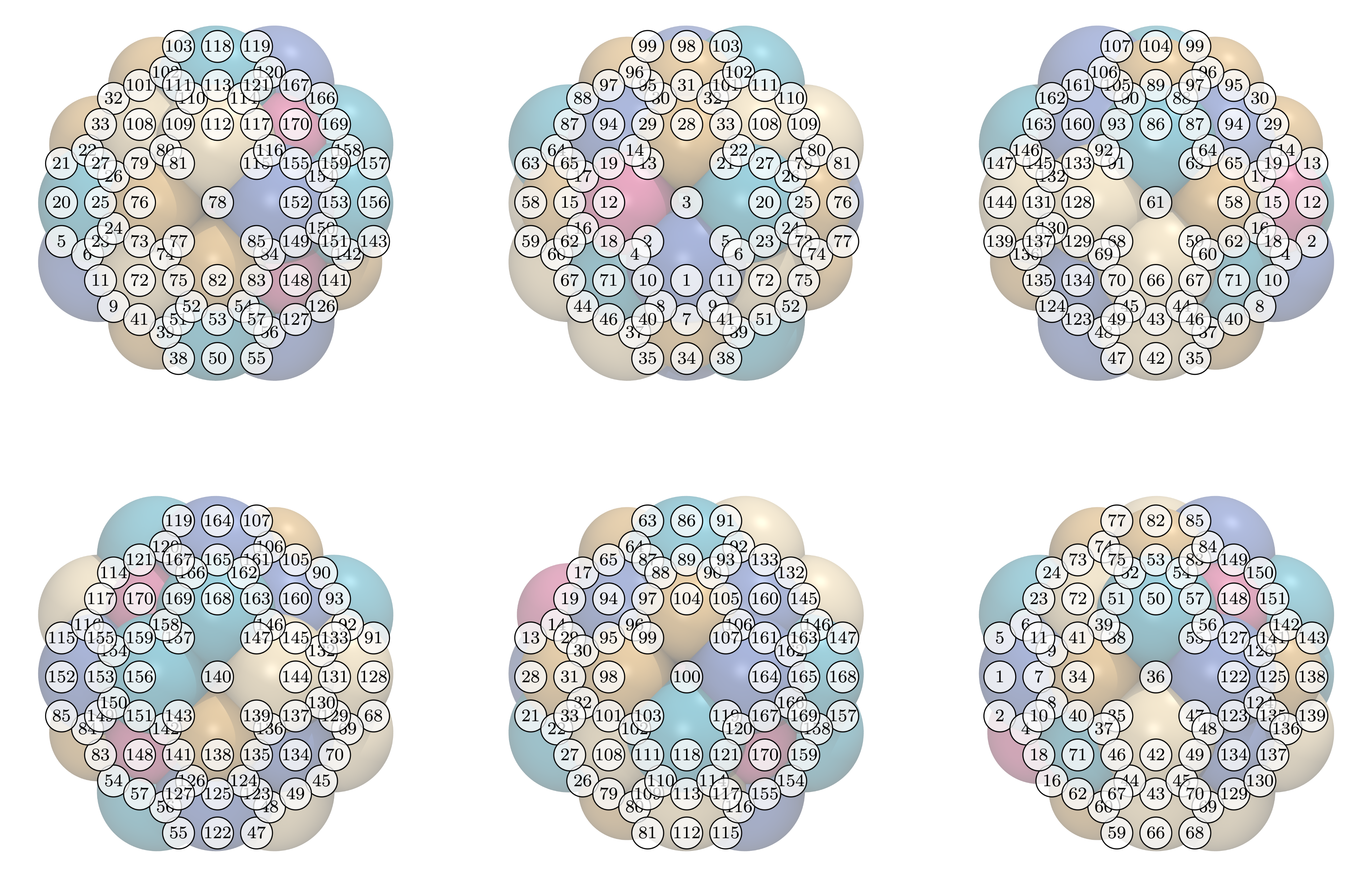}
    \centering
    \caption{\textbf{Enumeration of adsorption sites on the nanoparticle model.} All adsorption sites were identified and enumerated using ACAT. The indices for one of the 25 generated nanoparticles are shown from six different viewing angles.}
    \label{fig:Site_numbering_NP}
\end{figure*}

\begin{figure}
    \includegraphics{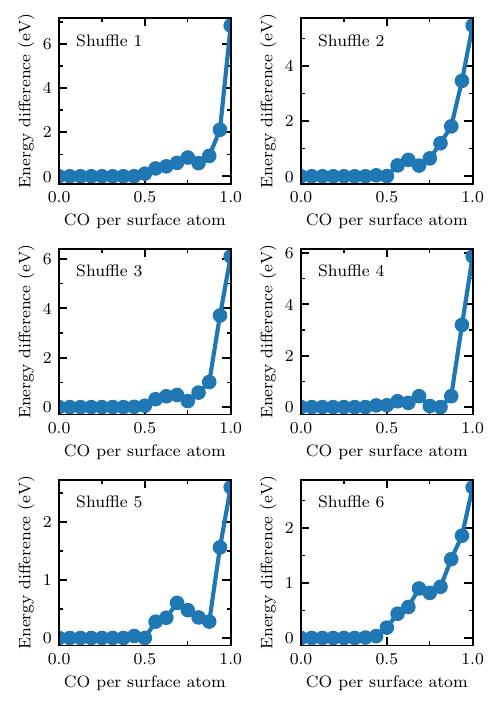}
    \centering
    \caption{\textbf{Difference in adsorption energy between configurations obtained using random and ACAT site enumeration.} The site indices shown in Fig.~\ref{fig:Site_numbering_slab} were shuffled six times. In each case, the configurations identified by TTOpt had higher adsorption energies than those obtained with the original enumeration provided by ACAT, once the surface coverage exceeds 0.5 CO per surface atom. Lines connecting the data points are drawn as guides for the eyes.}
    \label{fig:Shuffle_comparison}
\end{figure}

Fig.~\ref{fig:Site_numbering_slab} illustrates all 96 adsorption sites on the slab model along with their assigned indices $i$ provided by ACAT. Similarly, Fig.\ref{fig:Site_numbering_NP} shows the 170 adsorption sites on the nanoparticle model with their corresponding indices. The order in which the adsorption sites are enumerated is important, as these indices are used to map the occupation variables $\sigma_i$ to the corresponding sites. This, in turn, determines the structure of the QUBO coefficient matrix and HUBO coefficient tensor, which influences the performance of TTOpt for searching optimal adsorption configurations. 

When the indices are randomly assigned to the adsorption sites, the best adsorption configurations identified by TTOpt tend to be energetically higher than those obtained using the enumeration provided by ACAT. 
This is illustrated in Fig.~\ref{fig:Shuffle_comparison}, which shows the difference in adsorption energy between the configurations found using shuffled site indices and those obtained with the enumeration provided by ACAT for CO adsorption on the Pd–Pt slab. The site indices were randomly shuffled six times, and the results for each case are shown in the figure.
In all cases, once the surface coverage exceeds 0.5 CO per surface atom, the adsorption energies of the configurations identified by TTOpt are higher when the site indices are shuffled. Furthermore, the difference in adsorption energy increases as the surface coverage approaches 1 CO per surface atom.

We tested various renumbering schemes based on pairwise interaction energies between adsorbates and spatial distances between adsorption sites (data not shown). Since none of the tested schemes consistently led to configurations with lower adsorption energies, we used the enumeration provided by ACAT throughout this work.

\clearpage
\section*{References}
\bibliography{References.bib}

\end{document}